\documentclass[11pt,twoside]{article}  
\usepackage{adassconf}

\begin{document}   

%
%

\paperID{P.32}

%
%
%
%

\title{Specsim: The MIRI Medium Resolution Spectrometer Simulator}

%

\author{Nuria P.\ F.\ Lorente\altaffilmark{1,2}, Alistair C.\ H.\ Glasse 
and Gillian S.\ Wright}
\affil{UK Astronomy Technology Centre, Royal Observatory Edinburgh, 
Blackford Hill, Edinburgh, EH9 3HJ, United Kingdom}

\altaffiltext{1}{Email: npfl@roe.ac.uk}
\altaffiltext{2}{Previously Nuria P.\ F.\ M${\mathrm{ ^c}}$Kay}


\contact{Nuria P. F. Lorente}
\email{npfl@roe.ac.uk}

%
%
%
%
%

\paindex{Lorente, N. P. F.}
\aindex{Glasse, A. C. H.}     
\aindex{Wright, G. S.}     

%
%

\authormark{Lorente, Glasse \& Wright}




\begin{abstract}          
MIRI, the Mid-InfraRed Instrument, is one of four instruments being
built for the James Webb Space Telescope. It is being developed jointly
between an European Consortium (21 institutes from 10 countries, under
the auspices of ESA), and the US. MIRI consists of an imager, a
coronograph, a low-resolution spectrograph, and an Integral Field Unit
(IFU) Medium Resolution Spectrometer (MIRI-MRS). The latter will be the
first mid-infrared IFU spectrograph, and one of the first IFUs to be
used in a space mission. To give the MIRI community a preview of the
properties of the MIRI-MRS data products before the telescope is
operational, the Specsim tool has been developed to model, in software,
the operation of the spectrometer. Specsim generates synthetic data
frames approximating those which will be taken by the instrument in
orbit. The program models astronomical sources and transforms them into
detector frames using the predicted optical properties of the telescope
and MIRI. These frames can then be used to illustrate and inform a range
of operational activities, including data calibration strategies and the
development and testing of the data reduction software for the MIRI-MRS.
Specsim will serve as a means of communication between the many
consortium members by providing a way to easily illustrate the
performance of the spectrometer under different circumstances,
tolerances of components and design scenarios.
\end{abstract}


\section{The MIRI Medium Resolution Spectrometer}

The MIRI-MRS is an integral field unit (IFU)
spectrometer (Wright et al. 2004), which allows spectroscopy to be carried out
on a 2-dimensional area of sky in a single observation. The primary component of
the IFU is its image-slicing mirror, which divides the rectangular field of view
into a number of narrow slices. These are then arranged along the entrance slit
of a first-order diffraction grating, which carries out the dispersion.

The spectrometer operates over the range 5--28~$\mu$m, with a resolution of
${\mathrm{R}\sim3000}$. The spectral band is divided into 4 IFU channels, which
are observed simultaneously. Each channel is
equipped with an IFU image slicer designed to match the width of each slice to the
diffraction-limited point-spread function of the telescope, at the wavelength of
each channel. 

An observation over the entire spectral band is carried out in a set of three
exposures, the spectral band of each IFU channel being subdivided into 3
sub-bands by means of dichroic filters.
The data from each exposure are captured on one of two $1024\times1024$
pixel detectors (one for each pair of image slicer channels: (1A \& 2A), 
(3A \& 4A), (1B \& 2B) {\em etc}.
The expected sensitivity of the instrument is $1.2×10^{-20}$~Wm$^{-2}$ at
6.4~$\mu$m, and $5.6×10^{-20}$~Wm$^{-2}$ at 22.5~$\mu$m, and its field of view
widens with increasing channel number. The main functional parameters for the
MIRI-MRS are summarised in Table~\ref{tab:mrs_parameters}

\begin{deluxetable}{cccccc}
\label{tab:mrs_parameters}
\scriptsize
\tablecaption{Summary of the MIRI-MRS parameters}
\tablehead{
\colhead{Channel} & \colhead{FoV} & \colhead{Slices} & \colhead{$\lambda$} & 
\colhead{$\mathrm{R_{spectral}}$} & \colhead{Exposure}\\
\colhead{}        & \colhead{(arcsec$^2$)} & \colhead{} & \colhead{($\mu$m)} & \colhead{}& \colhead{}
}
\startdata
  &                  &     & 4.87 -- 5.82  & 2450 -- 3710 & A \nl
1 & 3.70$\times$3.70 & 21  & 5.62 -- 6.73  & 2450 -- 3710 & B \nl
  &                  &     & 6.49 -- 7.76  & 2450 -- 3710 & C \nl
\hline
  &                  &     & 7.45 -- 8.90  & 2480 -- 3690 & A \nl
2 & 4.70$\times$4.51 & 17  & 8.61 -- 10.28 & 2480 -- 3690 & B \nl
  &                  &     & 9.94 -- 11.87 & 2480 -- 3690 & C \nl
\hline
  &                  &     & 11.47 -- 13.67 & 2510 -- 3730 & A \nl
3 & 6.20$\times$6.13 & 16  & 13.25 -- 15.80 & 2510 -- 3730 & B \nl
  &                  &     & 15.30 -- 18.24 & 2510 -- 3730 & C \nl
\hline
  &                  &     & 17.54 -- 21.10 & 2070 -- 2490 & A \nl
4 & 7.74$\times$7.93 & 12  & 20.44 -- 24.72 & 2070 -- 2490 & B \nl
  &                  &     & 23.84 -- 28.82 & 2070 -- 2490 & C \nl
\enddata
\end{deluxetable}

The science goals of the MIRI-MRS encompass a broad area of study, including the
formation and evolution of galaxies, the life-cycle of stars and stellar systems,
the study of molecular clouds as the focus for star and planet formation, and
investigation of planetary evolution conducive to life (Gardener et al. 2005).

\section{Specsim: Modelling the Sky}

Specsim, an application developed in IDL, generates FITS frames which
approximate the images which will be produced by the MIRI-MRS.

The program allows the user to build a model of the instrument's field of view,
using a set of primitives to specify the morphological and spectral
characteristics of each astronomical target in the field. These currently
include point and extended sources, black-body and polynomial continuum
profiles, broad and narrow spectral lines and some commonly observed spectral
elements such as the 9.7 and 18$\mu$m silicate features. The target primitives
provided may be used in combination to construct model astronomical fields of
arbitrary complexity.

\begin{figure}
\plotone{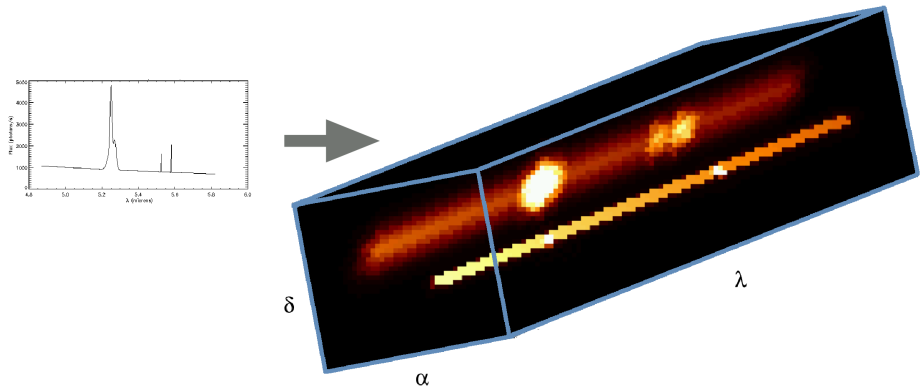}
\caption{Specsim target simulation cube (right), showing a point-source in continuum
with two narrow lines, and an extended source with a broad line and two narrow
spectral lines. The plot (left) shows the spectrum of the extended source,
generated by combining several of Specsim's target primitives. In
particular, the complex line at 5.25$\mu$m was constructed using three
broad-line components. The two narrow lines were modelled with a simple narrow
line function, and the continuum using the black-body target primitive.}
\label{fig:skycube}
\end{figure}

The user-specified target model, together with Specsim's internal models of the
zodiacal background, contributions from the telescope's solar shield, {\em
etc.}, are used to generate a target simulation cube, representing the
instrument's field of view over each channel's spectral range
(Figure~\ref{fig:skycube}). 

\begin{figure}
\plotone{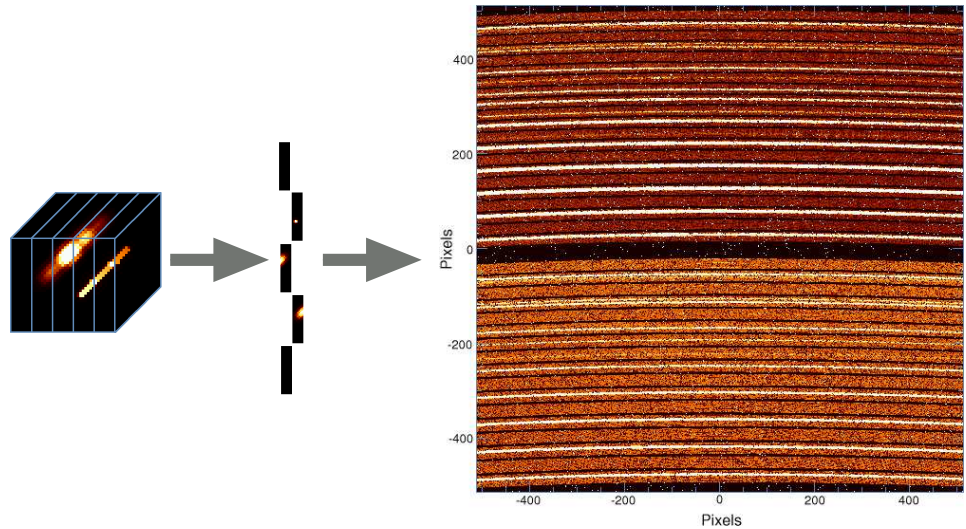}
\caption{Specsim segments the target cube to simulate the IFU image slicer (left),
the slices are aligned as for input into the diffraction grating (centre), and
finally they are dispersed. The simulated spectra for each pair of channels are
mapped onto each of two $1024\times1024$ detectors (right).}
\label{fig:detimg}
\end{figure}

\section{Specsim: Modelling the MIRI-MRS}
Once a model of the target field of view has been produced, Specsim simulates
the function of the spectrograph, producing a simulated spectroscopic
observation of the field.

This is carried out by first applying any contribution to the observed flux by
the instrument's optics and electronics to the sky model. Data files defining
these may be provided by the user, allowing the effect of instrument
characteristics (detector QE, optical efficiency, {\em etc}.) on the final
spectrometer image to be studied.
The next step in the process is to simulate the image slicer, 
by assigning each pixel in the sky model to an IFU slice, depending
on its location in the field of view. Geometric deformations can then be
applied, simulating the optical path of the telescope, the spectrometer's
pre-optics and the image-slicing mirror.
Dispersion of the image slices is then carried out, mapping these onto one half
of the detector image, according to IFU channel. Finally, cosmic rays,
photon and read-noise are added and on-sky integration is implemented 
(Figure~\ref{fig:detimg}).

Both the sky model cube and the detector image are provided as output for the
user. This allows the detector frames to be processed, analysed and compared
with the input targets, thus providing an useful test for the development of the
MIRI data reduction software, testing calibration strategies and
observation planning.

\vspace{1em}
\hspace{-1em}{\bf Acknowledgments.}
{\scriptsize MIRI draws on the expertise of the following organisations: Ames}
\parbox{\textwidth}{\scriptsize
Research Center, USA; Astron, Netherlands Foundation for Research in Astronomy;
CEA Service d'Astrophysique, Saclay, France; Centre Spatial de Li\'{e}ge,
Belgium; Consejo Superior de Investigacones Cient\'{i}ficas, Spain; Danish Space
Research Institute; Dublin Institute for Advanced Studies, Ireland; EADS
Astrium, Ltd., European Space Agency, Netherlands; UK; Institute d'Astrophysique
Spatiale, France; Instituto Nacional de T\'{e}cnica Aerospacial, Spain;
Institute of Astronomy, Zurich, Switzerland; Jet Propulsion Laboratory, USA;
Laboratoire d'Astrophysique de Marseille (LAM), France; Lockheed Advanced
Technology Center, USA; Max-Planck-Insitut f\"{u}r Astronomie (MPIA),
Heidelberg, Germany; Observatoire de Paris, France; Observatory of Geneva,
Switzerland; Paul Scherrer Institut, Switzerland; Physikalishes Institut, Bern,
Switzerland; Raytheon Vision Systems, USA; Rutherford Appleton Laboratory (RAL),
UK; Space Telescope Science Institute, USA; Toegepast-Natuurwetenschappelijk
Ondeszoek (TNO-TPD), Netherlands; UK Astronomy Technology Centre (UK-ATC);
University College, London, UK; Univ. of Amsterdam, Netherlands; Univ. of
Arizona, USA; Univ. of Cardiff, UK; Univ. of Cologne, Germany; Univ. of
Groningen, Netherlands; Univ. of Leicester, UK; Univ. of Leiden, Netherlands;
Univ. of Leuven, Belgium; Univ. of Stockholm, Sweden, Utah State Univ. USA 
}

\end{document}